\documentclass[twocolumn,showpacs]{revtex4}
\usepackage{graphicx}
\usepackage{dcolumn}
\usepackage{bm}
\usepackage{amsmath}
\usepackage{amsfonts}
\usepackage{amssymb}
\providecommand{\U}[1]{\protect\rule{.1in}{.1in}}

\begin{document}
\title{Generalized Jaynes-Cummings model as a quantum search algorithm }
\author{A. Romanelli}
\altaffiliation[{\textit{E-mail address:}}\\
]{alejo@fing.edu.uy}
\affiliation{Instituto de F\'{\i}sica, Facultad de Ingenier\'{\i}a\\
Universidad de la Rep\'ublica\\
C.C. 30, C.P. 11000, Montevideo, Uruguay}
\date{\today }

\begin{abstract}
%\vspace{0.2cm} \\
We propose a continuous time quantum search algorithm using a
generalization of the Jaynes-Cummings model. In this model the
states of the atom are the elements among which the algorithm
realizes the search, exciting resonances between the initial and the
searched states. This algorithm behaves like Grover's algorithm; the
optimal search time is proportional to the square root of the size
of the search set and the probability to find the searched state
oscillates periodically in time. In this frame, it is possible to
reinterpret the usual Jaynes-Cummings model as a trivial case of the
quantum search algorithm.
\end{abstract}

\maketitle

\section{Introduction}

One of the most simple and interesting quantum models that studies
the interaction between radiation and matter is the Jaynes-Cummings
model (JCM) \cite{Jaynes1}. The model considers the interaction
between a single two-level atom with a single mode of the
electromagnetic field. The coupling between the atom and the field
is characterized by a Rabi frequency, and a loss of excitation in
the atom appears as a gain in excitation of the field oscillator.
The collapse and the eventual revival of the Rabi oscillation,
described by the analytical solution of the JCM, shows a direct
evidence of the quantum nature of radiation. The use of JCM has
permitted to elucidate basic properties of quantum entanglement as
well as some aspects of the relationship between classical and
quantum physics. Since it was proposed, the pattern has been of
permanent interest in the quantum theory of interactions. In the
decade of the eighties it was discovered that the model exhibited
highly non classic behavior, and the possibility of experimental
realization appeared. The relative simplicity of the JCM and its
extensions has drawn much attention in the physics community and
recently in the field of the quantum computing \cite{Chuang,Azuma}.
In this work we use a generalization of the JCM to an $N$ state atom
interacting with a single field mode \cite{Jaynes2}.

In $1994$, Shor \cite{Shor} described a quantum algorithm to
decompose a number in its prime factors more efficiently than any
classic algorithm. It was exponentially faster than the best known
classical counterpart. In 2001 the experimental development of this
algorithm has had a very interesting advance: Vandersypen et
\emph{al.} \cite{Vandersypen} using a seven-qubit molecule
manipulated with nuclear magnetic resonance techniques has reported
the factorization of the number 15 into its prime factors 3 and 5.
This algorithm illustrates a part of the theoretical challenge of
quantum computation, \emph{i.e.} to learn how to work with quantum
properties to obtain more efficient algorithms. Tools like quantum
parallelism, unitary transformations, amplification techniques,
interference phenomena, quantum measurements, resonances, etc, must
be used by the new computation science. Grover, in $1997$, devised
an algorithm \cite{Grover} which can locate a marked item from an
unsorted list of $N$ items, in a number of steps proportional to
$\sqrt{N}$, that is quadratically faster than any classical
algorithm \cite{Boyer}. Continuous time search algorithms have been
investigated by a number of researchers \cite{Farhi,Childs,armando}.
The essential content of these proposals is to built a Hamiltonian
(or alternatively a unitary operator) with the aim to change the
initial average state $|\phi \rangle =\sum_{m}|\varphi _{m}\rangle
/\sqrt{N}$ into another state $|\varphi _{s}\rangle $, that also
belongs to the same set $\left\{ |\varphi
_{m}\rangle \right\} $ of $N$ vector in the base of the Hilbert space \cite%
{Chuang}. This last state is recognized by an unitary operator called
oracle, that is part of the global unitary evolution $exp(-iHt)$, where the
Hamiltonian is expressed as
\begin{equation}
H=|\phi \rangle \langle \phi |+|\varphi _{s}\rangle \langle \varphi _{s}|.
\label{hamiltonian1}
\end{equation}%
The probability to obtain the searched state is $\left\vert \langle \varphi
_{s}|exp(-iHt)|\phi \rangle \right\vert ^{2}$, and it equals $1$ after a
time $\pi \sqrt{N}/2$. In this frame, the search algorithm is seen as a
rotation in the Bloch sphere from the initial average state to the searched
state. Recently an alternative search algorithm was developed \cite%
{alejo,alejo1,alejo2,alejo3} that uses a Hamiltonian to produce a
resonance between the initial and the searched states, having the
same efficiency than the Grover algorithm. It can be implemented
using any Hamiltonian with a discrete energy spectrum, and it was
shown to be robust \cite{alejo} when the energy of the searched
state has some imprecision. The responses of this algorithm to an
external monochromatic field and to the decoherences introduced
through measurement processes was also analyzed in \cite{alejo1}.
However we do not know of any experimental implementation, not even
for a small search set. In this paper we present a resonant quantum
search algorithm implemented with a generalization of the two-level
JCM.

The paper is organized as follows. In section \ref{sec:generalizado}
we consider a generalization of the JCM, in section
\ref{sec:resonancia} we develop the search model. In Section
\ref{sec:Numerical} we present numerical results for our model.
Finally in Section \ref{sec:conclution} we draw some conclusions.

\section{Generalized Jaynes-Cummings model}
\label{sec:generalizado} We shall consider the generalization of the
JCM to an $N$ state atom interacting with a single field mode with
frequency $\omega $ synthesized by the following Hamiltonian
\cite{Jaynes2}
\begin{equation}
H=\hbar \omega \text{ }a^{\dagger }a+\sum\limits_{k=1}^{N}\varepsilon
_{k}S_{kk}+\frac{\hbar }{2}\Omega _{0}\text{ }\sum\limits_{k=1}^{N}\left(
a^{\dagger }S_{jk}+S_{kj}a\right) .  \label{hamilton}
\end{equation}%
The photon creation and annihilation operators $a^{\dagger }$and $a$
act on the photon number state $|n\rangle $ verifying
\begin{eqnarray}
\left[ a,a^{\dagger }\right] &=&1,  \label{commutador}\\
a^{\dagger }a|n\rangle &=&n|n\rangle ,\text{\ }  \label{prop1} \\
a^{\dagger }|n\rangle &=&\sqrt{n+1}|n+1\rangle ,  \label{prop2} \\
a|n\rangle &=&\sqrt{n}|n-1\rangle .  \label{prop3}
\end{eqnarray}%
$\varepsilon _{k}$ is the energy of atomic state $|\varphi _{k}\rangle $, $%
\Omega _{0}$ is the atom-field coupling constant and it is fixed by
physical considerations such as the cavity volume and the atomic
dipole moment, $S_{jk}$ is a transition operator acting on atomic
states defined by
\begin{equation}
S_{lk}|\varphi _{m}\rangle =\delta _{km}|\varphi _{l}\rangle ,  \label{rule}
\end{equation}%
where $\delta _{km}$ is the Kronecker delta. In what follows the subindex $j$
shall indicate the initial state of the atom $|\varphi _{j}\rangle $. This
state will be the starting state for the search algorithm and it can be
chosen as the ground state for experimental purposes.

Let us call $|\varphi _{s}\rangle $ the unknown searched state whose energy $%
\varepsilon _{s}$ is known. This knowledge is equivalent to ``mark"
the searched state in the Grover algorithm
\cite{Farhi,Childs,armando}. Our task
is to find the eigenvector $|\varphi _{s}\rangle $ with transition energy $%
\omega _{sj}=$ $\left( \varepsilon _{j}-\varepsilon _{s}\right) /\hbar $
from the given initial state $|\varphi _{j}\rangle $. Then it is necessary
to tune the frequency of the photon field with the frequency of the
transition $|\varphi _{j}\rangle \longrightarrow |\varphi _{s}\rangle $.
This means that the frequency of the cavity mode is selected as $\omega
\equiv \omega _{sj}$. The transition between the atomic states is governed
by the interaction term of the Hamiltonian Eq.(\ref{hamilton})
\begin{equation}
W=\frac{\Omega _{0}}{2}\text{ }\sum\limits_{k=1}^{N}\left( a^{\dagger
}S_{jk}+S_{kj}a\right) .  \label{transition}
\end{equation}%
The transition probability to pass from the initial atomic state
$|\varphi _{j}\rangle $ with $m$ photons to any final state
$|\varphi _{i}\rangle $ with $n$ photons is proportional to
$\left\vert \langle n\varphi _{i}|W|m\varphi _{j}\rangle \right\vert
^{2}$. After some steps we get
\begin{equation}
\left\vert \langle n\varphi _{i}|W|m\varphi _{j}\rangle \right\vert ^{2}=%
\frac{\Omega _{0}^{2}}{4}\left\{ (m+1)\text{ }\delta _{nm+1}+m\text{ }\delta
_{nm-1}\text{ }\delta _{ij}\right\} .  \label{transition1}
\end{equation}%
To calculate this transition probability independently of the
initial and final numbers of photons the statistical weight of the
photons must be incorporated,
\begin{equation}
P_{ji}=\frac{1}{\lambda ^{2}}\sum\limits_{n}\sum\limits_{m}p(n)p(m)\text{ }%
\left\vert \langle n\varphi _{i}|W|m\varphi _{j}\rangle \right\vert ^{2},
\label{conjunta}
\end{equation}%
where $p(n)$ is the normalized photon number distribution and
$\lambda$ is an unknown constant. Using Eq.(\ref{transition1}) in
Eq.(\ref{conjunta}) we obtain the dependence of $P_{ji}$ with the
average number of photons and the parameter $\Omega _{0}$,
\begin{equation}
P_{ji}=\frac{1}{\lambda ^{2}}\frac{\Omega _{0}^{2}}{4}\left\{ \left\langle
n\right\rangle +1+\delta _{ij}\left\langle n\right\rangle \right\} ,
\label{conjunta2}
\end{equation}%
where $\left\langle n\right\rangle =\sum\limits_{n}n$ $p(n)$. Taking
into account the normalization condition
$\sum\limits_{i=1}^{N}P_{ji}=1$, the dependence $\Omega _{0}$ with
the number of atom levels is obtained,
\begin{equation}
\Omega _{0}=\frac{2\lambda }{\sqrt{\left\langle n\right\rangle \left(
N+1\right) +N}}\sim \frac{2\lambda }{\sqrt{\left\langle n\right\rangle N}},
\label{omega}
\end{equation}%
The last step is valid for large $N$ and $\left\langle n\right\rangle $.
Note that $\Omega _{0}$ depends on the photon distribution function only
through the mean value of $n$.

\section{Resonances}

\label{sec:resonancia} In the previous section we have determined
the dependence of the atom-field coupling constant $\Omega _{0}$
with the number of atomic states $N$ and the mean number of photons
$\left\langle n\right\rangle $. Now we want to study how this
coupling constant determines the characteristic period of the
dynamics and subsequently the waiting time for the search algorithm.

The dynamics of the system is given by the the Schr\"{o}dinger
equation for the wavefunction $|\Psi (t)\rangle $,
\begin{equation}
i\hbar \frac{\partial |\Psi (t)\rangle }{\partial t}=H|\Psi (t)\rangle ,
\label{Schrodinger}
\end{equation}%
where $H$ is given by Eq.(\ref{hamilton}). The global Hilbert space
of the system is built as the tensor product of the spaces for the
photons and the atom. Therefore atom-field wavefunction $|\Psi
(t)\rangle $ is expressed as a linear combination of the basis
$\left\{ \left\vert \varphi _{m}\right\rangle |n\rangle \right\} $
\begin{equation}
|\Psi (t)\rangle =\sum_{m}\sum_{n}b_{mn}\exp \left[ -i\left(
\varepsilon _{m}+\omega n\right) t/\hbar \right] \left\vert \varphi
_{m}\right\rangle \left\vert n\right\rangle ,  \label{psi1}
\end{equation}%
where $\left\{ |n\rangle \right\} $ is the basis for the photons and
$\left\{ \left\vert \varphi _{m}\right\rangle \right\} $ is the
eigenvector basis for the atomic Hamiltonian without electromagnetic
field. The phase factor is introduced to simplify the final
differential equations. Substituting Eq.(\ref{psi1}) into
Eq.(\ref{Schrodinger}) and projecting on the state $\left\langle
\varphi _{l}\right\vert \left\langle k\right\vert $ the following
set of differential equations for the time depended amplitudes
$b_{lk}(t)$ are obtained
\begin{equation}
\frac{2i}{\Omega _{0}}\frac{db_{lk}}{dt}=\sqrt{k}\exp \left[ -i\text{ }%
\left( \omega _{lj}-\omega _{sj}\right) \text{ }t/\hbar \right] b_{jk-1}
\notag
\end{equation}%
\begin{equation}
+\delta _{lj}\sqrt{k+1}\sum_{m=1}^{N}b_{mk+1}\exp \left[ -i\left( \omega
_{lm}+\omega _{sj}\right) t/\hbar \right] .  \label{dife}
\end{equation}%
This set of equations shall be solved numerically in the next
section. Here we proceed to study their qualitative behavior. These
equations have two time scales involved, a fast scale associated to
the Bohr frequencies $\omega _{lk}$, and a slow scale associated to
the amplitudes $b_{lk}(t)$. If we are interested in the slow scale
all terms that have fast phase in the previous equations can be
ignored; the most important terms are the ones with zero phase. In
this approximation the previous set of equations Eq.(\ref{dife})
becomes
\begin{equation}
\frac{2i}{\Omega _{0}}\frac{db_{sk}}{dt}=\sqrt{k}b_{jk-1}  \label{apro1}
\end{equation}%
\begin{equation}
\frac{2i}{\Omega _{0}}\frac{db_{jk}}{dt}\simeq \sqrt{k+1}b_{sk+1}.
\label{apro2}
\end{equation}%
These equations represent two oscillators that are coupled so that
their population probabilities alternate in time. As we notice the
coupling is established between the initial and the searched for
states. To uncouple the previous equations we combine them to obtain
\begin{equation}
\frac{d^{2}b_{jk}}{dt^{2}}\simeq -{(k+1)}\frac{\text{ }\Omega _{0}^{2}}{4}%
\text{ }b_{jk},  \label{inicial}
\end{equation}%
\begin{equation}
\frac{d^{2}b_{sk}}{dt^{2}}\simeq -k\frac{\text{ }\Omega _{0}^{2}}{4}\text{ }%
b_{sk}.  \label{search}
\end{equation}%
Solving these equations, for a given number $k$ of photons and with
the initial conditions $b_{jk}(0)=1$ and $b_{sk}(0)=0$, the
following results for the amplitudes are obtained
\begin{equation}
b_{jk}(t)\simeq \cos (\frac{\Omega _{0}}{2}\sqrt{k+1}t),
\label{binicial}
\end{equation}%
\begin{equation}
b_{sk+1}(t)\simeq -i\sin (\frac{\Omega _{0}}{2}\sqrt{k+1}t).
\label{bbuscado}
\end{equation}%
Therefore the probabilities to obtain the initial and the searched
states, independently of the initial number of photons, should be
calculated as $P_{j}(t)=\sum\limits_{k}p(k)\left\vert
b_{jk}(t)\right\vert ^{2}$, $P_{s}(t)=\sum\limits_{k}p(k)\left\vert
b_{sk}(t)\right\vert ^{2}$, which satisfy the conditions
$P_{j}(0)=1$, $P_{s}(0)=0$. Averaging over the number of photons and
using Eq.(\ref{omega}) these probabilities are
\begin{equation}
P_{j}(t)\simeq \cos ^{2}\left( \Omega \text{ }t\right) ,  \label{pj}
\end{equation}%
\begin{equation}
P_{s}(t)\simeq \sin ^{2}\left( \Omega \text{ }t\right) ,  \label{ps}
\end{equation}
where the new angular frequency is
\begin{equation}
\Omega =\frac{\lambda} {\sqrt{N}}. \label{omega0}
\end{equation}%
Then we see that the probabilities of the initial state $%
P_{j}(t) $ and the searched state $P_{s}(t)$ oscillate harmonically
with a frequency $\Omega $ and period
$T=\frac{2\pi}{\lambda}\sqrt{N}$, while the probability of the other
elements of the search set are neglected. If we let the system
evolve during a time
\begin{equation}
\tau \equiv \frac{T}{4}=\frac{\pi }{2\lambda }\sqrt{N},
\label{periodo}
\end{equation}%
and at this precise moment we make a measurement, we have
probability $1$ to obtain the searched state. It is important to
indicate that this approach is valid only in the adiabatic
approximation \cite{schiff}, this means that all the
 frequencies $\omega _{nm}$ are much larger than $\Omega _{0} $.
Therefore the efficiency of our search algorithm is the same as that
of the Grover algorithm and additionally it is independent of the
number of photons.

In the next section, we implement numerically Eq.(\ref{dife}) and we
show that this agrees with the above theoretical developments of the
system in resonance.

\section{Numerical results}
\label{sec:Numerical} The JCM has been used to understand the
behavior of circular Rydeberg atoms, where the valence electron is
confined near the classical Bohr orbit \cite{Brune3}. This suggests
to choose for our purpose an atomic model with an attractive
potential whose quantum energy eigenvalues are $\varepsilon
_{n}=-\varepsilon _{0}/n^{2}$, where $n$ \ is the principal quantum
number and $\varepsilon _{0}$ is a parameter. In this frame the Bohr
transition frequency can be expressed as function of the parameter
$\varepsilon _{0}$ as $\omega _{nm}=\varepsilon
_{0}(1/n^{2}-1/m^{2}) $.
\begin{figure}[th]
\begin{center}
\includegraphics[scale=0.38]{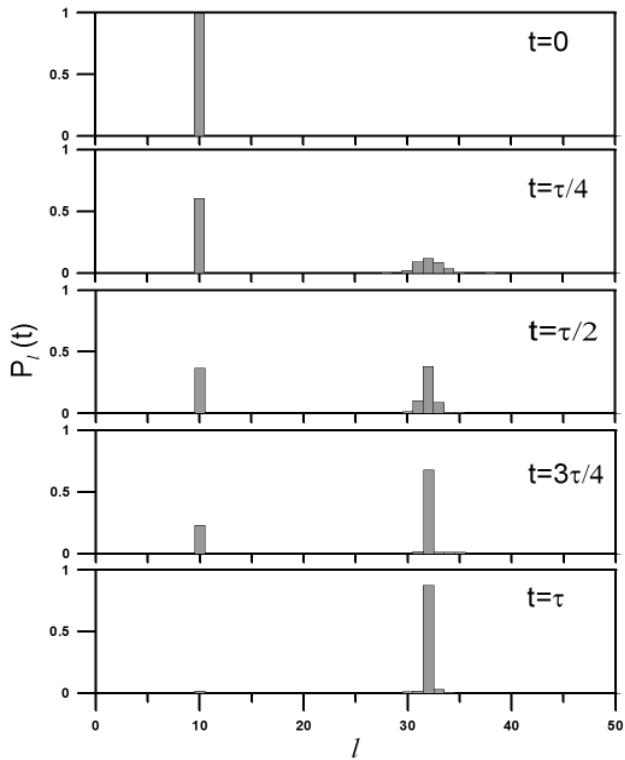}
\end{center}
\caption{Probability distribution for all atomic levels at five
different times. The search set $\left\{l\right\}$, with
$l=1,2,...,N$, has $N=50$ elements (levels), $\delta=40000$ and
$\tau$ is the optimal search time proportional to $\sqrt{N}$. The
initial state was taken to be $j=10$ and the searched state $s=32$.
Note that the distribution probability is essentially shared between
the initial and the searched states.} \label{t2}
\end{figure}

\begin{figure}[th]
\begin{center}
\includegraphics[scale=0.38]{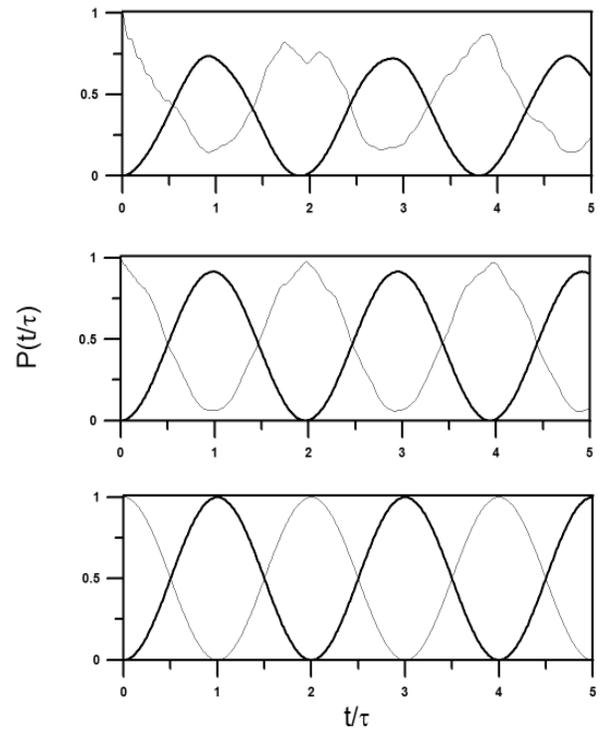}
\end{center}
\caption{Probability distributions as functions of time for the
initial (thin line) and the searched (thick line) states. The
parameter $\delta$ takes the values, from top to bottom, $5\times
10^3$, $10^4$ and $10^6$. The size of the search set is $N=50$.}
\label{t1}
\end{figure}

We shall choose the numerical values of the parameters taking into
account some previous experimental data. In Ref.
\cite{Brune3,Randall} the Rabi oscillation of circular Rydberg atoms
was observed. The frequency of the single electromagnetic mode was
tuned with the transition between adjacent circular Rydberg states
with principal
quantum numbers $51$ and $50$, where the fast scale associated to this Bohr frequency was $%
\omega _{50,51}\thicksim100\pi$ GHz, that is very large compared
with the fundamental Rabi frequency $\Omega _{0}\thicksim 50\pi$
KHz. In the micromaser configuration of Ref. \cite{Rempe} the field
frequency of $21$ GHz produced the Rydberg transitions used in this
experiment, where the principal quantum number was about $63$ and
the Rabi frequency $10$ KHz. From the above we conclude that the
ratio between the Bohr transition energy and the vacuum Rabi
frequency $\delta \equiv \varepsilon _{0}/\Omega _{0}$ to be used in
our search algorithm should be taken as $\sim 10^{6}$.

We have integrated numerically Eq.(\ref{dife}) varying the parameter
$\delta$ in a range between $5\times 10^{3}$ and $10^{6}$. The
initial conditions are (i) a uniform distribution for the photons
and (ii) $b_{jk}(0)$ is chosen in such a way that $P_{j}(0)=1 $. The
calculations were performed using a standard fourth order
Runge-Kutta algorithm. The procedure consisted in choosing at random
the energy of the searched state and then to follow the dynamics of
the set and of the initial state. We verified for several values of
$\delta$ that the most important coupling is between the initial and
the searched states; the other couplings may be totally neglected.
\newline

In Fig.\ref{t2} we show the probabilities for all levels at five
different fractions of the time $\tau$. We see that the dispersion
among the states neighboring the searched state is relatively small
for $\delta=40000$. For higher values of $\delta$ the dispersion is
even smaller, this confirms that our theoretical approximation of
two coupled modes is correct. Furthermore we conclude that the flux
transfer process is essentially an interchange between the initial
and the searched states and that the optimal time to measure the
searched state is $\tau$. At other times we have less chance of
perform a measurement of the searched state.

Fig.\ref{t1} shows the oscillation of the probability flux between
the initial and the searched states as a function of time for three
values of $\delta$. The time is normalized for the theoretical
characteristic time $\tau$. The evolution shows for the lower values
of $\delta$ an almost periodic behavior, however for the highest
value $\delta=10^6$ the behavior is completely harmonic. In this
last case there is clearly a characteristic time when the
probability of the searched state is maximum and very near $1$ and
the initial state probability is near $0$. The optimal time agrees
with our theoretical prediction $\tau$. This periodic behavior and
the proportionality between $\tau$ and $\sqrt N$ are also found in
the Grover algorithm \cite{Chuang}.

\section{Conclusions}

\label{sec:conclution}

In this work we show how a generalized JCM to an $N$ state atom
interacting with a single field mode can be thought of as a quantum
search algorithm that behaves like the Grover algorithm; in
particular the optimal search time is proportional to the square
root of the size of the search set and the probability to find the
searched state oscillates periodically in time.

In the past, the biggest difficulty to build a JCM has been to
obtain a single electromagnetic mode that interacts with the atomic
transition. This difficulty has been overcome in the last decades
thanks to the experimental advances in the handling of Rydberg atoms
and to the building of micro-cavity for microwaves
\cite{Brune1,Brune2,Brune3,Randall,Rempe}. In the frame of this work
we can interpret these devices as experimental realizations of the
``analog" \cite{Farhi} Grover algorithm in the trivial case of the
search of a marked item in an unsorted list of 2 elements. However
this new way of looking at the problem is different from the usual
point of view and opens new possibilities for the JCM.  In summary,
in this paper we reinterpret the JCM as the first step to build a
more generic search algorithm with a large number of elements

\bigskip

I thank Eugenio Rold\'an for discussions that helped motivate this
work and Victor Micenmacher for comments. This work was supported by
PEDECIBA and ANII.

\end{document}